\documentclass[9pt, conference]{IEEEtran}

\IEEEoverridecommandlockouts
\usepackage{array}
\usepackage{cite}
\usepackage{amsmath,amssymb,amsfonts}
\usepackage{algorithmic}
\usepackage{graphicx}
\usepackage{textcomp}
\usepackage{xcolor}
\usepackage{booktabs}
\usepackage{multirow}
\usepackage[numbers]{natbib} 
\usepackage{url}
\usepackage{hyperref}

\def\BibTeX{{\rm B\kern-.05em{\sc i\kern-.025em b}\kern-.08em
    T\kern-.1667em\lower.7ex\hbox{E}\kern-.125emX}}
\begin{document}

\title{Text2FX: Harnessing CLAP Embeddings for Text-Guided Audio Effects \\
\thanks{This work was supported by NSF Award Number 2222369.}
}


\author{\IEEEauthorblockN{Annie Chu}
\IEEEauthorblockA{
\textit{Northwestern University}}
\and
\IEEEauthorblockN{Patrick O'Reilly}
\IEEEauthorblockA{
\textit{Northwestern University}}
\and
\IEEEauthorblockN{Julia Barnett}
\IEEEauthorblockA{
\textit{Northwestern University}}
\and
\IEEEauthorblockN{Bryan Pardo}
\IEEEauthorblockA{
\textit{Northwestern University}}
}
\maketitle

\begin{abstract}
This work introduces Text2FX, a method that leverages CLAP embeddings and differentiable digital signal processing to control audio effects, such as equalization and reverberation, using open-vocabulary natural language prompts (e.g., ``make this sound in-your-face and bold''). Text2FX operates without retraining any models, relying instead on single-instance optimization within the existing embedding space, thus enabling a flexible, scalable approach to open-vocabulary sound transformations through interpretable and disentangled FX manipulation. We show that CLAP encodes valuable information for controlling audio effects and propose two optimization approaches using CLAP to map text to audio effect parameters. While we demonstrate with CLAP, this approach is applicable to any shared text-audio embedding space. Similarly, while we demonstrate with equalization and reverberation, any differentiable audio effect may be controlled.  
We conduct a listener study with diverse text prompts and source audio to evaluate the quality and alignment of these methods with human perception. Demos and code are available at \url{anniejchu.github.io/text2fx}

\end{abstract}


\section{Introduction}


Audio effects (e.g., equalization, reverberation, compression) are essential tools in modern audio production. From mainstream pop to podcasts to film scores, audio effects (FX) are integral in shaping the final sound. However, their complex and often unintuitive controls (e.g., decay, cutoff frequency) can be extremely challenging for non-experts and time-consuming for professionals. For instance, despite its seemingly straightforward description, transforming a simple drum recording into the `crunchy hyperpop' drum sound of Charli XCX may require a complex process involving the careful adjustment of over 20 distinct effect parameters across multiple FX, such as distortion, saturation, equalization, and compression.

Semantic audio production research aims to bridge the gap between \textit{high-level concepts} (e.g., `old time telephone') and \textit{signal-level effect parameters} (e.g., controls of a parametric equalizer) \cite{moffat2022semantic}. Pre-deep-learning efforts, such as Sabin et al. \cite{sabin2011weighted} and Audealize \cite{seetharaman2016audealize}, used crowdsourcing to map natural language terms to specific effect parameters, such as equalization (EQ) or reverberation (Reverb). While effective, these methods produced closed-vocabulary mappings limited to single FX, unable to generalize beyond new words or phrases. This work also resulted in  word-parameter setting datasets for single FX, such as SocialFX \cite{zheng2016socialfx} (EQ, Reverb, compression) 
and SAFE \cite{stables2016semantic} (four open-source plugins). Most recently, Balasubramaniam et al. \cite{balasubramaniam2023word} explored text-driven audio manipulation by training a deep model on the EQ subset of Audealize \cite{seetharaman2016audealize}. However, as their approach focuses on text-to-audio generation rather than directly mapping text to effect parameters, it functions as a black box, limiting users' ability to shape the final result. Like earlier work, it is limited by the closed vocabulary of single-word descriptors from training. We seek to overcome these limitations by exploring method that enables open-vocabulary text prompts to control any set of differentiable effects without retraining for new words or FX. 

\begin{figure}
\centerline{\includegraphics[width=9cm] 
{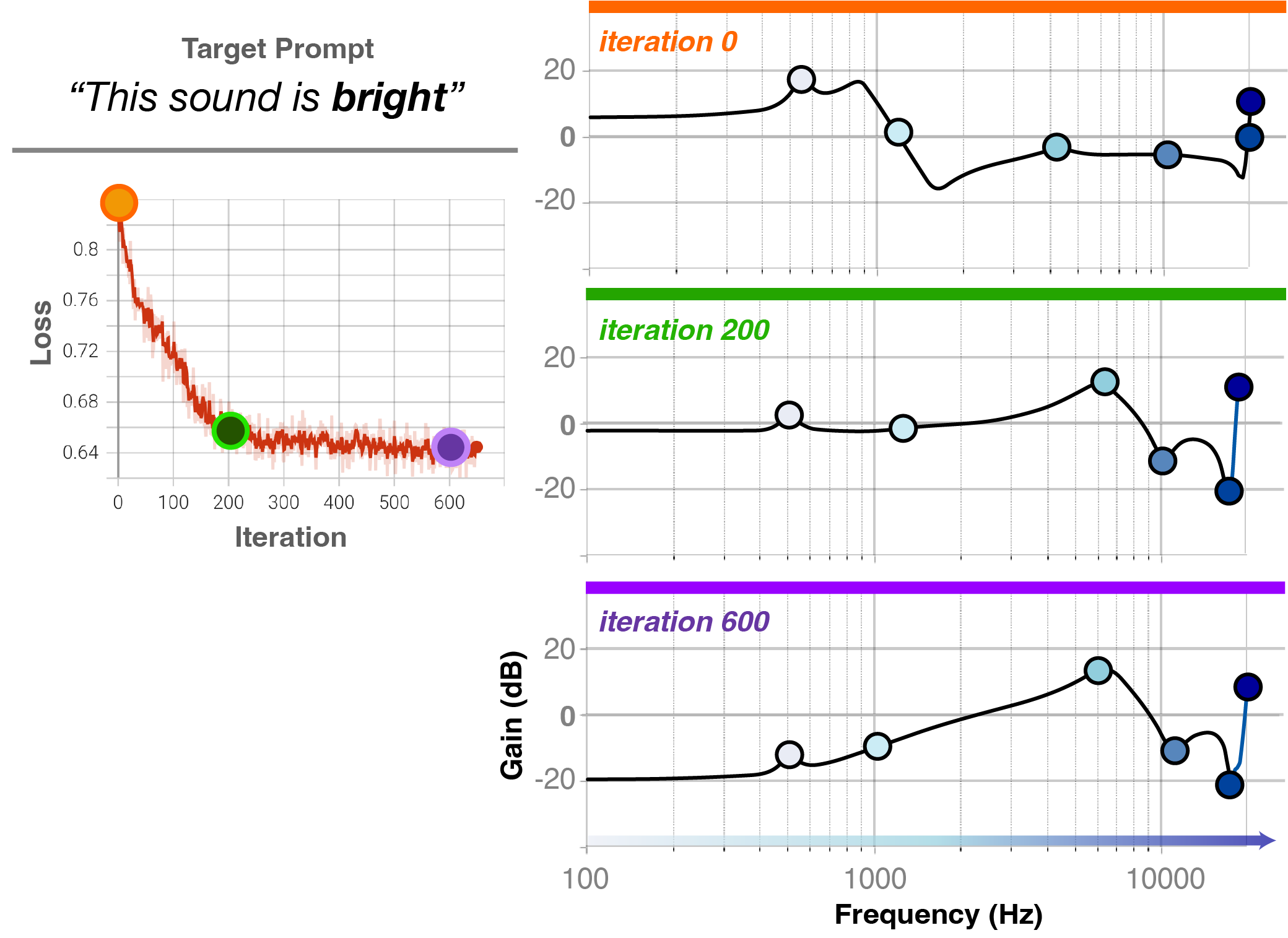}}
\caption{\textbf{Text2FX Example.} A previous study \cite{seetharaman2016audealize} found listeners associate `bright' with boosting high frequencies ($>$ 2 kHz) and cutting low ones ($<$ 2 kHz). Optimizing the audio in a shared text-audio embedding space (CLAP) towards the embedding for text `bright' achieves this;. \textbf{Left:} Optimization loss curve. \textbf{Right:} Estimated settings for a 6-band  parametric EQ.}
\label{optimization example}
\end{figure}
Recent large multimodal embedding models like CLAP \cite{elizalde2023clap} have made great strides in bridging natural language with audio. Trained on a diverse, extensive dataset of paired audio-text captions, CLAP features 
a joint embedding space aligning audio with corresponding textual descriptions. 
Though successfully applied to zero-shot classification and audio captioning \cite{elizalde2023clap}, as well as text-to-audio generation \cite{cherep2024creativetexttoaudiogenerationsynthesizer}, CLAP's ability to encode qualitative notions of audio FX---such as what constitutes a `bright' sound--- remains unexplored.  

Differentiable digital signal processing (DDSP) \cite{engel2020ddsp, hayes2024review} allows traditional DSP parameters (e.g., filter coefficients, gain controls, and synthesis parameters) to be learned through gradient-based optimization. DDSP has been successfully applied in tasks including speech synthesis \cite{fabbro2020speech}, synthesizer-based sound generation \cite{cherep2024creativetexttoaudiogenerationsynthesizer}, style transfer for audio FX \cite{steinmetz2022styletransferaudioeffects}, and mastering \cite{vanka2024diffmstdifferentiablemixingstyle}, but has not been applied to text-driven audio FX. 

In this paper, we explore whether CLAP embeddings contain actionable knowledge for natural language-based control of audio FX. To leverage this knowledge, we introduce Text2FX, a method that uses CLAP's learned representations to manipulate audio FX through cross-modal optimization. Integrating CLAP with DDSP, Text2FX performs single-instance optimization within the audio FX parameter space, aligning the audio embedding with that of a given text description. Given an audio recording, a prompt (e.g., `shrill and sharp'), and an FX chain (i.e., sequence of audio FX like EQ $\rightarrow$ Reverb), Text2FX generates both the ``effected'' audio along with the interpretable, adjustable FX parameters applied to achieve the desired effect. We aim to lay the foundation for a future intuitive open-vocabulary natural language interface, allowing users to hear the ``effected'' audio and further refine the ballpark FX parameters generated by the system. Given the subjective nature of semantic descriptors (e.g., `warm'), it is essential the system returns FX parameters users can adjust to suit their individual preferences. Our goal is not to replace expert knowledge but to ease the learning curve for beginners and inspire creativity. Our contributions are as follows:
\begin{enumerate} 
    \item We demonstrate that CLAP embeddings contain relevant 
    information for open-vocabulary control of audio FX.  
    \item We propose two single-instance optimization approaches harnessing CLAP to apply and tune audio FX (EQ and Reverb) with open-vocabulary natural language prompts. 
    \item We perform a listener study to assess the quality and alignment of these approaches with listener expectations. 
\end{enumerate}

\section{Proposed Method: Single-Instance Optimization via CLAP Tuning}
\subsection{Method Overview}
In our method, Text2FX, we start by selecting a target prompt that describes the desired outcome (e.g., `bright'). We then apply randomly-selected parameter settings to each effect in the designated FX chain (e.g., EQ $\rightarrow$ Reverb) and process the audio with these parameters (FXparams). This ``effected'' audio and target prompt are passed through CLAP to generate embeddings in the shared text-audio embedding space.
We then perform single-instance optimization by iteratively adjusting the FXparams through gradient-based optimization such that the embedding of the ``effected'' audio gets closer to the desired position in the embedding space. At the end of the optimization, the resulting FXparams are open to inspection and modification by the user.

This method optimizes examples within an existing embedding space (CLAP) to identify suitable parameters in the FX parameter space, repurposing an off-the-shelf embedding model that has never been trained for audio effect applications, a technique similar to Tagbox \cite{manilow2022steering}. No additional model training and no additional models are used. Crucially, the system’s vocabulary is encoded by CLAP, which was trained on the 4.6M audio-text pairs of Audioset \cite{gemmeke2017audio}, far surpassing the small datasets of single-word descriptors relied on by all prior work. This ensures strong adaptability and generalizability. While we pair CLAP with EQ and Reverb as the FX, the approach is easily adaptable to any model with a shared text-audio embedding space and any set of differentiably implemented audio FX. Also, while we focus on differentiable effects, the method can in theory be extended to non-differentiable FX by replacing gradient descent with a gradient-free optimizer  \cite{cherep2024creativetexttoaudiogenerationsynthesizer, steinmetz2024stitocontrollingaudioeffects}.



\subsection{Two optimization approaches}
\begin{figure}
\centerline{\includegraphics[width=9cm] 
{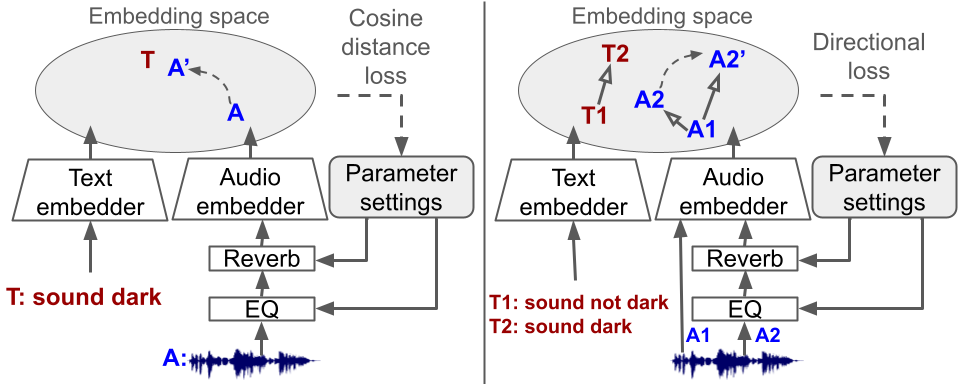}}
\caption{\textbf{Left, Text2FX-cosine:} Input audio (A) and target prompt (T) are mapped into the same (CLAP) embedding space. A is optimized to move its embedding closer to T, resulting in modified audio (A'). \textbf{Right, Text2FX-directional:} Both the directional vector between a contrasting prompt (T1) and target prompt (T2) and the vector between input audio (A1) and `effected' audio (A2) are measured. A2 is optimized to make the vector between audio embeddings align with the vector between text embeddings, resulting in A2'.}
\label{fig1}
\end{figure}

When repurposing CLAP embeddings to guide the application of audio FX, a question arises: Should the ``effected'' audio embedding be made as similar as possible to the target text prompt embedding, or should it follow the directional change between embeddings of two contrasting text prompts (e.g., `not warm' $\rightarrow$ `warm')?  To answer this question, we compare two approaches, illustrated in Figure \ref{fig1}.

The first approach, \textit{Text2FX-cosine}, aims to minimize the cosine distance between the embedding of the ``effected'' audio and that of the fixed text target, 
directly moving the audio embedding towards the text embedding. 
While straightforward, we hypothesize this approach may lead to unintended consequences. For example, if the input is already `warm' and the target is to ``make it warm'', the optimization may result in no change (as the audio is already `warm') or shift focus towards altering the audio content (such as increasing volume) rather than enhancing its quality (warmth).

To mitigate this, \textit{Text2FX-directional} leverages the directional relationship between two embedding pairs. This method guides the ``effected'' audio embedding to move in the direction defined by the difference between the text target and a contrasting text prompt (see Figure \ref{fig1}). This approach, initially proposed for CLIP embeddings \cite{radford2021CLIP} for image editing \cite{kim2022diffusionclip}, is adapted here for CLAP. 

In \textit{Text2FX-directional}, we generate four embeddings: A1 (the fixed starting audio), A2 (the optimized ``effected" audio), T1 (the contrasting text prompt; e.g., `NOT warm'), and T2 (the target text prompt; e.g., `warm'). With these, a guiding direction aligned with the desired audio transformation is specified, facilitating the optimization process to steer the audio embedding along the direction of the intended change, rather than simply moving it closer to the target text embedding.
This approach operates under the implicit assumption that if the user is asking to make a sound `warm', they believe the starting audio is \textit{\textbf{not} warm}. As per CLAP's training methodology \cite{elizalde2023clap}, we prepend a phrase ``this sound is'' to the text before embedding.


\subsection{Optimization Details}
We use a learning rate of 1e-2, the Adam optimizer, and apply a random shift of at most 1500ms to the audio signal at every iteration to prevent model fixation on audio content (e.g., a distorted \textit{guitar riff}) and encourage focus on audio quality (e.g., a \textit{distorted} guitar riff). We initialize FXparams randomly from a standard normal distribution. 
As preliminary experiments showed convergence within 300-400 iterations, final optimization was extended to 600 iterations to ensure thorough convergence. To account for the stochastic nature of random initialization, we perform three runs for each instance and select the run with the lowest loss, doing this for both variants.

\section{Empirical Validation: Listener Study}
We conducted a listener study to evaluate how the outputs produced by Text2FX (both variants) align with human expectations. 
\subsection{Preparing Evaluations}
\begin{table*}[h!]
\begin{center}
\caption{Natural Language Prompts}
\begin{tabular}{>{\raggedright\arraybackslash}p{0.9cm}>{\raggedright\arraybackslash}p{2.0cm}>{\raggedright\arraybackslash}p{1.4cm}>{\raggedright\arraybackslash}p{4.3cm}>{\raggedright\arraybackslash}p{7.2cm}}
\toprule
\textbf{} & \multicolumn{2}{c}{\textbf{Single Words (10)}} & \multicolumn{2}{c}{\textbf{Multiwords (10)}} \\
\cmidrule(r){2-3} \cmidrule(r){4-5}
\textbf{} & \textbf{\textit{Concrete (7)}} & \textbf{\textit{Abstract (3)}} & \textbf{\textit{Combination (5)}} & \textbf{\textit{Imagery (5)}} \\
\midrule
\multirow{3}{*}{\textbf{EQ}} & tinny, muffled, light, deep, crisp, bright, mellow & ethereal, eerie, grand & soft yet vibrant, in-your-face and bold, shrill and sharp, quiet and gentle, cool and smooth & coming through an old telephone, coming from a speaker under a blanket, booming like a thunderstorm, delivered with a softer feel, like a hazy surreal dream \\
\midrule
\multirow{5}{*}{\textbf{Reverb}} & boomy, spacious, dry, cavernous, echoey, underwater,  reverberant & empty, long, bold & booming and vast, clear but distant, cozy and enveloping, heavy and dramatic, hollow and far-away & coming from a cathedral, coming from a long hallway, coming from a small and intimate sound booth, like an explosion in a canyon, accompanied by a faint atmospheric haze in the background \\
\midrule
 & metallic, harsh, & 
dramatic, & barren and detached, warm and &
coming from a small cavern with a muffled echo, coming \\
\textbf{EQ $\rightarrow$} & cold, blaring, & 
fluffy, & 
full-bodied, vibrant and powerful, &
from underwater in a swimming pool, coming from a broken\\
\textbf{Reverb} & 
bassy, grainy, & 
powerful & 
resonant and harmonious, high and &
speaker in an empty warehouse, like a shrill Victorian ghost, \\
&
breezy & 
& 
tinny &
like a distant radio broadcast with a warm lingering presence \\
\bottomrule
\label{tab-nlp-prompts}
\end{tabular}
\end{center}
\vspace{-3mm}
\end{table*}

\textbf{FX Chains:}
We evaluated on two FX, EQ and Reverb, across three distinct FX chain configurations: 1) EQ-only, 2) Reverb-only, and 3) EQ $\rightarrow$ Reverb. We chose these FX due to their widespread use and the availability of a semantic audio dataset, Audealize \cite{seetharaman2016audealize}, that provides human-validated text labels and effect settings. This dataset also informed the selection of single-word descriptors for our natural language prompts. 
The EQ and Reverb used in our experiments are the standard 6-band Parametric EQ (18 parameters) and NoiseShapedReverb (23 parameters) from the dasp\footnote{github.com/csteinmetz1/dasp-pytorch} library. 

\textbf{Natural Language Prompts:}
We selected 20 natural language text prompts for each FX chain, totaling 60 unique prompts. Each FX chain has a distinct set of prompts (see Table \ref{tab-nlp-prompts}). Prompts were categorized into two main groups: 10 single-word prompts and 10 multi-word prompts per FX chain. We aimed for a diverse set of prompts, covering a wide range of perceptual attributes, contextual descriptions, and levels of semantic concreteness (e.g., `bright' is more tangible and concrete than `hopeful'). Single-word prompts were sourced directly from the Audealize \cite{seetharaman2016audealize} dataset as it provides a set of high-quality, human-validated terms for EQ and Reverb. 
For EQ-only and Reverb-only FX chains, prompts were drawn from the corresponding Audealize subsets. For the EQ → Reverb FX chain, prompts were selected from the overlap of both subsets. Using Audealize's agreement metric, we selected a majority of high-confidence, \textit{concrete} words (e.g., `warm') and a few low-agreement, \textit{abstract} words (e.g., `happy'). 
As Audealize only provides single-word descriptors, multi-word prompts were crafted by combining and expanding these terms. These included both straightforward combinations (e.g., `light and airy') as well as more evocative, imagery-based descriptions (e.g., `from a speaker under a blanket'). 

\textbf{Audio Stimuli:}\label{subsec-audio-stimuli}
We curated a diverse set of 30 reference audio recordings (15 speech, 15 music) from public datasets, including MusDB18 \cite{musdb18}, VocalSet \cite{wilkins2018vocalset}, IDMT-SMT-GUITAR \cite{kehling2014automatic}, daps \cite{mysore2014can}, and LibriTTS \cite{zen2019libritts}. This selection includes various instruments (mono and polyphonic), gender-balanced speech, and diverse acoustic environments. For each reference, four ``effected'' outputs were generated:

\begin{itemize}
    \item \textbf{Text2FX-cosine:} FXparams optimized via cosine loss 
    \item \textbf{Text2FX-directional:} FXparams optimized via directional loss
    \item \textbf{Random:} Randomly assigned FXparams
    \item \textbf{noFX:} The original reference audio without any FX
\end{itemize}

The noFX version served as a baseline and additional attention check to ensure reliability of participants' evaluations. We applied each natural language prompt to 4 audio files (2 music, 2 speech), resulting in 80 unique text-audio sets per FX chain. Each set was assessed by 5 participants, totaling 1200 evaluations across the three FX chains (EQ-only, Reverb-only, EQ → Reverb), with each prompt evaluated 20 times.

\subsection{Participant Task and Inclusion Criteria}
Through Prolific, we recruited 200 English-speaking adults who completed 100+ tasks with an approval score of $\geq$ 95\% to complete all 1200 evaluations, with each participant completing 6 evaluations. Participants underwent a listening screening from Rumbold et al. \cite{rumbold2024correlations} to measure sensitivity to tones from 55 Hz to 10 kHz, and we measured music engagement using Zhang et al.'s \cite{zhang2019single} single-question predictor. Individuals who failed the listening test or self-identified as tone-deaf were excluded due to insufficient audio sensitivity.
We additionally discarded evaluations where the noFX sample was rated outside the range of [-0.5, 0.5] on the evaluation scale to address unreliable data from rushed participants, particularly in later evaluations. Following data cleaning, the dataset includes 167 participants and 924 evaluations. 

An evaluation consisted of a target prompt (e.g. `warm'), the original reference audio, and the 4 ``effected'' versions of the same audio, processed as described in Section \ref{subsec-audio-stimuli}. Participants were given instructions to evaluate how much more or less `warm,' for example, each processed audio sounded relative to the original. We provided a continuous rating scale of -2 to 2 labeled as follows:
\begin{itemize}
    \item \textbf{+2:} The audio changed in the right direction (i.e., definitely more \textit{warm} than reference)
    \item \textbf{0:} No noticeable change compared to the reference (neutral)
    \item \textbf{-2:} The audio changed in the wrong or unrelated direction (i.e., changed, but definitely not more \textit{warm} than reference)
\end{itemize}

\section{Experimental Results}
\begin{figure*}[ht!]
\centerline{\includegraphics[width=18.5cm] 
{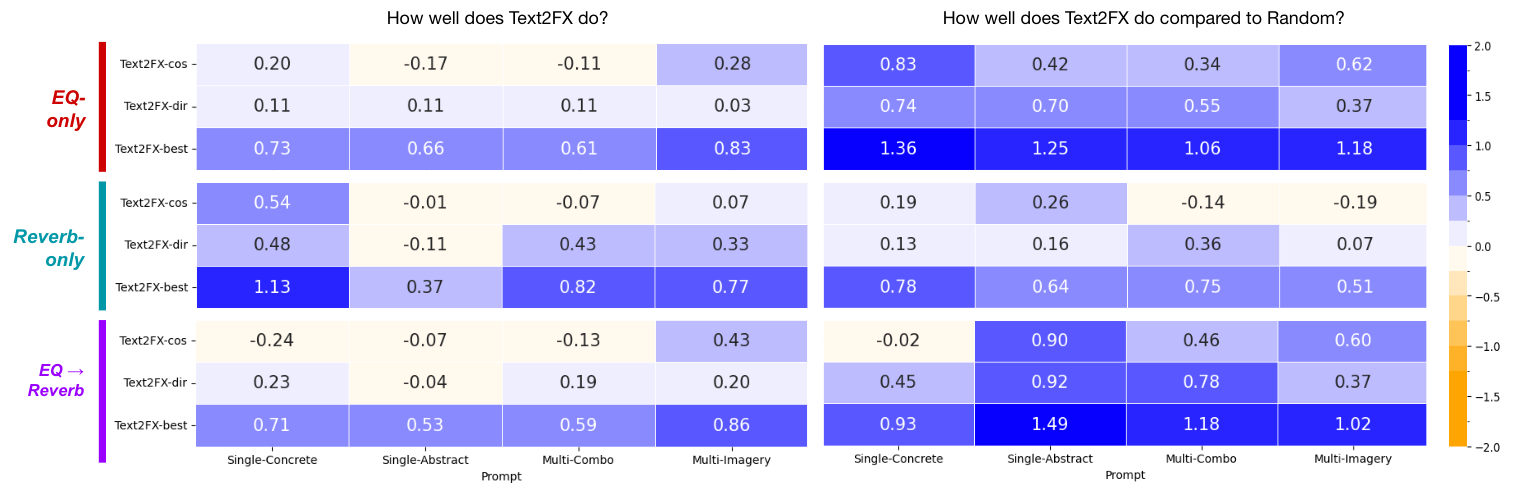}}
\caption{\textbf{Left:} The mean listener evaluation score. \textbf{Right:} The amount by which the mean evaluation score beats the mean listener evaluation score achieved by a random effect. Higher numbers are better. In all conditions, Text2FX-best has a positive mean listener score and always beats Random.}
\label{heatmap}
\end{figure*}

\begin{table}[b]
\begin{center}
\vspace{-4mm}
\caption{Percentage (\%) of Evaluations Resulting in Positive Scores}
\begin{tabular}{lccc}
\toprule
\textbf{Model} & \textbf{EQ} & \textbf{Reverb} & \textbf{EQ $\rightarrow$ Reverb} \\
\midrule
Text2FX-cosine      & 48.26 & 51.61  & 47.24\\
Text2FX-directional & 45.49 & 53.23 & 50.61 \\
Random             & 22.22 & 49.03 & 30.37  \\
Text2FX-Best      & 67.01 & 74.19 & 68.10 \\
Text2FX-Both        & 26.74 & 30.65 & 29.75 \\
\bottomrule
\end{tabular}
\label{tab3}
\end{center}
\end{table}
\subsection{Does CLAP contain useful knowledge for audio FX control?}
To address this question in a falsifiable manner, we reformulate as: For each approach, what percentage of the 924 listener evaluations resulted in positive ratings? We also measure the percentage of evaluations where the higher-scoring optimization approach received a positive score (Text2FX-Best) and where both approaches received positive scores (Text2FX-Both).

In Table \ref{tab3}, we see both Text2FX variants outperform Random, achieving a success rate of about 50\%. Text2FX-Best shows success in 67-74\% of cases, indicating at least one variant effectively achieved the target prompt for the large majority of word-audio combinations. Pearson correlation analysis reveals negative correlations between the listener evaluations of the two Text2FX variants: -0.25 for EQ-only, -0.22 for Reverb-only, and -0.24 for EQ → Reverb, suggesting the two variants have distinct strengths and excel in different contexts. 

A more granular analysis is presented in Figure \ref{heatmap}. Figure \ref{heatmap} (left) displays the mean listener evaluation scores across all prompt categories and FX chains. While successful in all subcategories, \textbf{the best-performing Text2FX variant exhibited particularly strong performance with single-concrete and multi-imagery prompts.} This is consistent with the nature of concrete and imagery words, which are closely tied to physical properties or objects, aligning well with CLAP’s training on AudioSet \cite{gemmeke2017audio}—a dataset of audio clips annotated with sound event labels.
 
Figure \ref{heatmap} (right) displays the mean difference in listener evaluation scores between the method in question and the random baseline. It reveals a substantial and consistent performance gap between Text2FX, when applying the best-performing variant, and Random, particularly pronounced for EQ and EQ → Reverb FX chains, with differences in ratings consistently exceeding 1.0 on the original listener scale of -2 to 2. 

We conclude \textbf{CLAP does encode relevant information for controlling audio FX}, with our findings suggesting the effectiveness may vary depending on text prompt characteristics. Given this, the primary challenge becomes optimizing CLAP's application to better suit the diversity of text prompts and their corresponding audio FX.

\subsection{What is the best way to leverage CLAP?}
To understand the strengths and weaknesses of each optimization variant, we investigate their performance across the different prompt subcategories. As shown in Figure \ref{heatmap}, \textbf{Text2FX-directional consistently achieves the target transformation with greater reliability than Text2FX-cosine.} Text2FX-directional achieved the target transformation in 10 of 12 prompt categories, while Text2FX-cosine succeeded in only 5. The two cases in which Text2FX-directional struggled were single-abstract prompts in Reverb-only and EQ $\rightarrow$ Reverb.
Interestingly, while Text2FX-directional consistently delivers more subtle-to-moderate changes in the desired direction, \textbf{Text2FX-cosine produces more polarizing transformations}—performing exceptionally well 
in some cases (e.g., single-concrete for EQ-only) but very poorly in others (e.g., single-concrete for EQ $\rightarrow$ Reverb). Finally, although Text2FX-cosine occasionally outperforms Text2FX-directional in EQ-only and Reverb-only for single-word prompts, Text2FX-directional consistently performs better with multi-word prompts and the longer EQ $\rightarrow$ Reverb FX chain. For EQ $\rightarrow$ Reverb, Text2FX-directional surpasses Text2FX-cosine in all cases except multi-imagery. This suggests \textbf{a directional loss function is better suited for generalizing to longer prompts and complex FX chains}.

\section{Conclusions}
This work introduces Text2FX, a method integrating CLAP with DDSP to control audio effects through natural language descriptions. Advancing semantic audio production, Text2FX opens new possibilities for educational tools and creative explorations in audio effects, enabling intuitive and customizable audio manipulation via open-vocabulary text prompts. We highlight the following findings:

\begin{enumerate}
    \item \textbf{CLAP encodes relevant information for audio FX:} At least one CLAP optimization improves listener ratings in 67-75\% of cases, indicating meaningful qualitative encoding of audio FX transformations. While it performs best on prompts related to physical objects, it can also use abstract and combined prompts.
    \item \textbf{Both proposed optimization approaches work:} Text2FX-cosine and Text2FX-directional produce distinct outputs and listeners vary in which output they prefer, depending on the prompt-audio-FX chain configuration. This is advantageous, as each variant can compensate for the other's limitations. 
    \item \textbf{Directional loss shows greater potential for generalization:} Text2FX-directional generally outperforms Text2FX-cosine, though the latter produces better results in some cases. Our study suggests Text2Fx-directional is better able to to generalize to longer prompts and FX chains, suggesting directional loss is a promising avenue for further research.
\end{enumerate}

Future research may seek to explore a broader range of language prompts and instruction-based controls similar to InstructPix2Pix \cite{brooks2023instructpix2pixlearningfollowimage}, investigate more complex FX chain configurations, and develop an interactive human-in-the-loop interface. 

\bibliographystyle{IEEEtran}
\bibliography{IEEEabrv,references}
\end{document}